\documentclass[preprints,article,accept,pdftex,moreauthors,bioengineering,pdftex]{Definitions/mdpi} 
\usepackage{xcolor}
\usepackage{multirow}

\definecolor{blue1}{RGB}{79, 113, 190}
\definecolor{orange1}{RGB}{239, 137, 51}
\definecolor{graynew}{gray}{0.88}
\definecolor{purple1}{RGB}{112, 48, 160}

\newcommand{\blue}[1]{\textcolor{blue1}{#1}}
\newcommand{\orange}[1]{\textcolor{orange1}{#1}}
\newcommand{\purple}[1]{\textcolor{purple1}{#1}}

\firstpage{1} 
\pubvolume{1}
\issuenum{1}
\articlenumber{0}
\pubyear{2022}
\copyrightyear{2022}
\datereceived{} 
\dateaccepted{} 
\datepublished{} 
\hreflink{https://doi.org/} 

\Title{High-fidelity Direct Contrast Synthesis from Magnetic Resonance Fingerprinting}

\TitleCitation{Title}

\Author{Ke Wang $^{1}$*\orcidA{}, Mariya Doneva $^{2}$, Jakob Meineke $^{2}$, Thomas Amthor $^{2}$, Ekin Karasan $^{1}$, Fei Tan $^{3}$, Jonathan I. Tamir $^{4}$, Stella X. Yu $^{1,5,6}$ and Michael Lustig $^{1}$}

\AuthorNames{Ke Wang, Mariya Doneva, Jakob Meineke, Thomas Amthor, Ekin Karasan, Fei Tan, Jonathan I. Tamir, Stella X. Yu, and Michael Lustig}

\AuthorCitation{Wang, K.; Doneva, M.; Meineke, J; Amthor, T.; Karasan E.; Tan, F.; Tamir, J.; Yu, S; Lustig, M}

\address{%
$^{1}$ \quad Electrical Engineering and Computer Sciences, University of California, Berkeley, Berkeley, CA, USA\\
$^{2}$ \quad Philips Research Europe, Hamburg, Germany\\
$^{3}$ \quad Bioengineering, UC Berkeley-UCSF, San Francisco, CA, USA\\
$^{4}$ \quad Chandra Family Department of Electrical and Computer Engineering, The University of Texas at Austin, Austin, TX, USA\\
$^{5}$ \quad International Computer Science Institute, Berkeley, CA, USA\\
$^{6}$ \quad Computer Science and Engineering, University of Michigan, Ann Arbor, MI, USA\\
}

\corres{Correspondence: kewang@berkeley.edu}

\abstract{Magnetic Resonance Fingerprinting (MRF) is an efficient quantitative MRI technique that can extract important tissue and system parameters such as T1, T2, B0, and B1 from a single scan. This property also makes it attractive for retrospectively synthesizing contrast-weighted images. 
In general, contrast-weighted images like T1-weighted, T2-weighted, etc. can be synthesized directly from parameter maps through spin-dynamics simulation (\emph{i.e.}, Bloch or Extended Phase Graph models). However, these approaches often exhibit artifacts due to imperfections in the mapping, the sequence modeling, and the data acquisition. 
Here we propose a supervised learning-based method that directly synthesizes contrast-weighted images from the MRF data without going through the quantitative mapping and spin-dynamics simulation.
To implement our direct contrast synthesis (DCS) method, we deploy a conditional Generative Adversarial Network (GAN) framework and propose a multi-branch U-Net as the generator. The input MRF data are used to directly synthesize T1-weighted, T2-weighted, and fluid-attenuated inversion recovery (FLAIR) images through supervised training on paired MRF and target spin echo-based contrast-weighted scans. 
\emph{In-vivo} experiments demonstrate excellent image quality compared to simulation-based contrast synthesis and previous DCS methods, both visually as well as by quantitative metrics. 
We also demonstrate cases where our trained model is able to mitigate in-flow and spiral off-resonance artifacts that are typically seen in MRF reconstructions and thus more faithfully represent conventional spin echo-based contrast-weighted images.
} 
\begin{document}


\section{Introduction \label{sec:introduction}}

Magnetic resonance imaging (MRI) is an effective imaging modality offering tremendous benefits to both science and medicine.
The main advantage of MRI is the richness of soft tissue contrasts that can be generated by simply changing the pulse sequence parameters. Image contrast in MRI is dominated by biophysical tissue properties such as Proton Density (PD), longitudinal/transverse relaxation (T1/T2), magnetic susceptibility,  and diffusion, to name a few. These parameters provide information on the tissue composition and its micro-structure and are excellent biomarkers for diagnosing and assessing disease. 
Measuring the quantitative value of tissue parameters, \emph {i.e.}, quantitative MRI (qMRI), is desirable as it could provide a standardized metric of tissue property~\cite{pierpaoli2010quantitative}. However, qMRI has been notoriously challenging to implement and standardize in clinical practice. Traditional mapping sequences require many lengthy scans to map a single parameter, thus unsuitable for rapid imaging. Consequently, today's diagnostic exams are composed of a series of several scans, each \emph{qualitatively} emphasizing one of the physical parameters above. For example, the routine brain MRI exam includes PD-weighted scans where brighter pixel intensities indicate a higher density of protons, T1-weighted (T1w) scans where brighter intensities indicate shorter T1 recovery, T2-weighted (T2w) scans where brightness indicates longer T2 relaxation, Fluid-attenuated inversion recovery (T1/T2-FLAIR) where fluid signals are suppressed, and diffusion scans in which brighter intensities indicate less diffusivity. The relative contrast differences within and across these scans can aid in the assessment of disease.

Owing to the need for multiple scans to obtain multiple contrasts, the typical MRI protocol is lengthy, requiring patients to keep still for tens of minutes and hindering scanner throughput.
In recent years, there have been notable research efforts in acquiring or synthesizing multi-contrast images from single or fewer scans to shorten the total exam time ~\cite{tanenbaum2017synthetic,tamir2017t2,wang2021outcomes,ma2018fast,ma2013magnetic,wang2019echo}, with initial early success in clinical practice ~\cite{hsieh2020magnetic,vargas2016synthetic}.
For example, Synthetic MR methods~\cite{blystad2012synthetic,granberg2016clinical,tanenbaum2017synthetic,warntjes2008rapid,hsieh2020magnetic} acquire multiple short scans and use parameter fitting and physical models to simulate a variety of contrast-weighted images.
T2 Shuffling~\cite{tamir2017t2,tamir2019targeted} reconstructs multiple contrast-weighted images along the transverse relaxation curve using a single volumetric fast spin echo (FSE) acquisition by randomly shuffling the phase encode view ordering and performing subspace modeling.
Similarly, multitasking~\cite{christodoulou2018magnetic,cao2022three} approaches use tensor low-rank constraints to reconstruct multiple contrast-weighted images from a single rapid acquisition. Common to the above approaches is the careful choice of scan parameters to reduce confounding factors and isolate a small number of qMRI parameters contributing to the overall image contrast.
%
%


Instead of reducing confounding factors, an alternative approach known as Magnetic Resonance Fingerprinting (MRF)~\cite{ma2013magnetic,ma2018fast,hsieh2020magnetic} proposed to mix many quantitative parameters using a short acquisition with randomized scan parameters. MRF has accelerated the pace of clinical qMRI by demonstrating the ability to rapidly and reliably generate multiple quantitative parameter maps from a single scan.
An MRF acquisition is usually based on gradient echo sequences and consists of rapid repetition times (TR) with under-sampled spiral readouts, where the flip angle is modified every TR such that the steady state of spin dynamic is never achieved. MRF produces a sequence of images in which tissue with different relaxation and field properties (T1, T2, PD, B0, B1) will produce a unique time series or "fingerprint". 
The quantitative parameters of the tissue are then extracted by matching the resulting time series of each pixel to the closest signal in a  precomputed dictionary constructed by simulating the Bloch equation for parameter combinations within a realistic range. 

The ability to estimate quantitative parameters from a single sequence means that the information to synthesize contrast-weighted images is also embedded in MRF. And while quantitative parameter maps provide meaningful physical parameters of tissue, clinicians still primarily rely on contrast-weighted images for clinical diagnosis. This opens the opportunity for MRF to be a one-stop shop sequence that provides both parameter maps and synthetic contrast MRI. 

\begin{figure}[!ht]
\hspace{-35mm}
\includegraphics[width=1.25\linewidth]{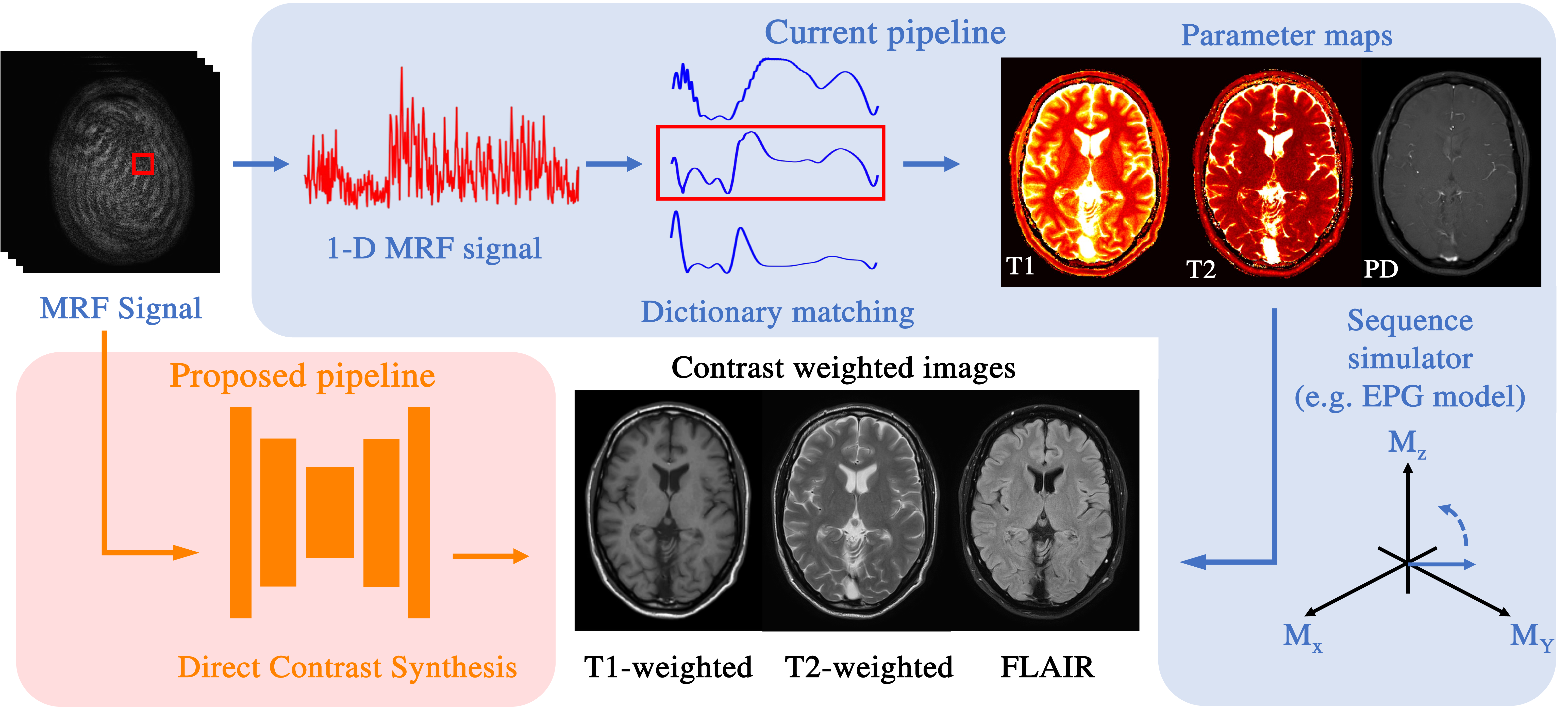}
\caption{\label{fig:mrfoverview}\textbf{Contrast synthesis from MRF via \blue{current simulation-based pipeline} and proposed \orange{direct contrast synthesis (DCS) pipeline}.} Simulation-based method takes the predicted quantitative parameter maps from MRF and synthesizes different contrast-weighted images by simulating the MRI physics. Our proposed DCS uses a spatial CNN to transform the MRF time series directly into different contrast-weighted images. DCS bypasses dictionary matching and contrast simulation steps, avoids modeling and acquisition imperfections, and produces high-fidelity contrast-weighted images.}
\end{figure}

\begin{figure}[!ht]
\hspace{-30mm}
\includegraphics[width=1.25\linewidth]{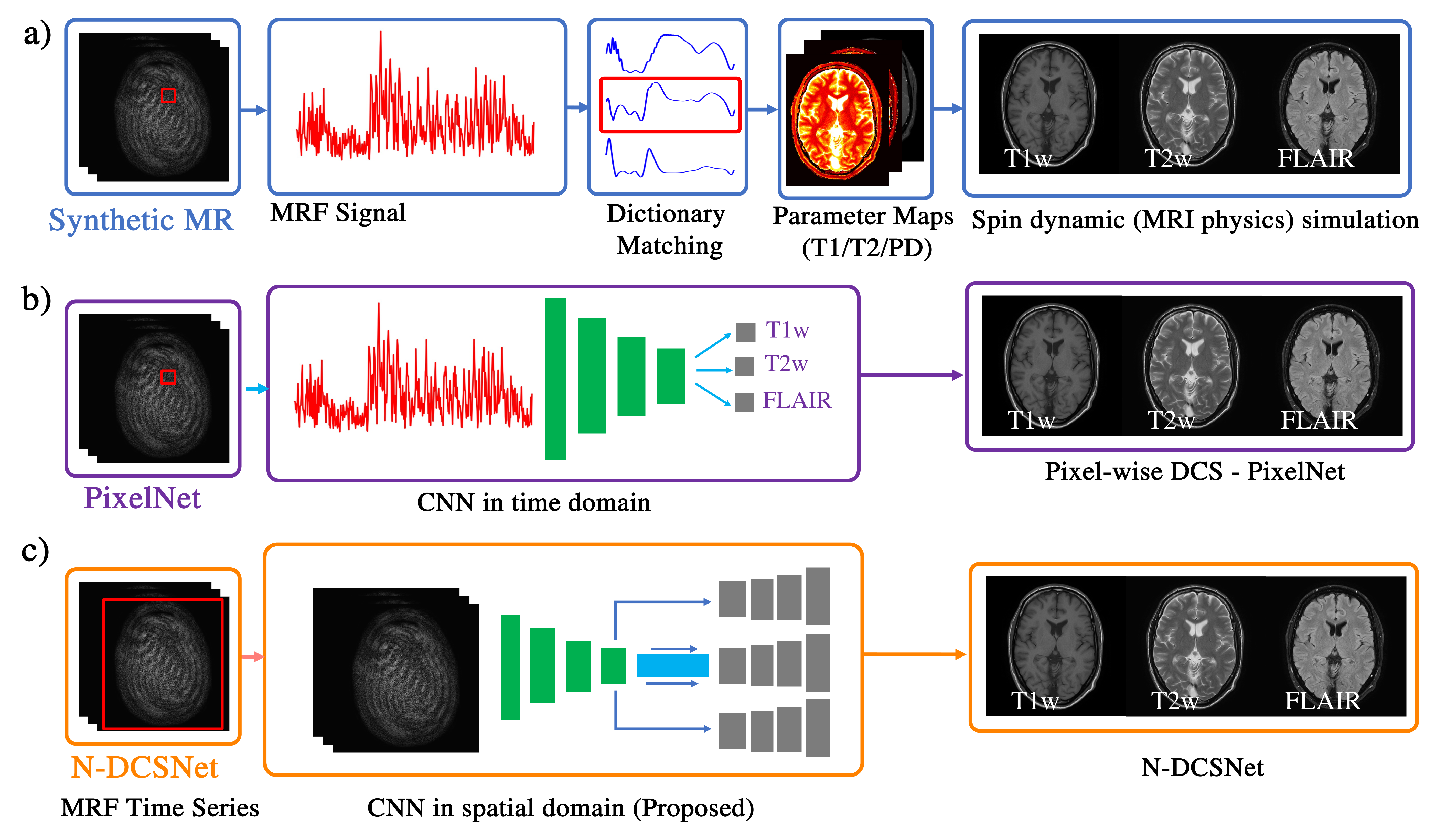}
\caption{\label{fig:dcscomparisons}\textbf{Three possible pipelines to generate contrast-weighted images from MRF.} a) \textbf{\blue{Synthetic MR}} generates multi-contrast images through dictionary matching and sequence simulation (e.g., Bloch equation, EPG). b) \textbf{\purple{PixelNet}} uses a 1-D pixel-wise time-domain CNN to output a qualitative contrast weighting for each voxel. c) Proposed \textbf{\orange{N-DCSNet}} leverages a GAN-based architecture and spatial-convolutional network to synthesize multi-contrast images. }
\end{figure}

One approach to synthesizing contrast-weighted images from MRF is to first fit the quantitative parameters and then simulate the contrast-weighted images~\cite{blystad2012synthetic}.
Figure~\ref{fig:mrfoverview} and Figure~\ref{fig:dcscomparisons}a show the spin-dynamic simulation pipeline, which takes quantitative parameter maps and uses them to synthesize different contrast-weighted images using the Bloch equation or Extended Phase Graphs (EPG)~\cite{weigel2015extended}.
Unfortunately, contrast-weighted images generated this way often exhibit artifacts due to many sources of error. Errors could arise from discrepancies between the MRF sequence and the dictionary simulation, for example, when flow, diffusion, magnetization transfer, excitation slice profile, or partial volume is not modeled appropriately. This limitation is most notable in FLAIR contrast, where errors are seen along the boundaries of cerebrospinal fluid (CSF)~\cite{vargas2016synthetic}.

An alternative and relatively more straightforward pipeline is to avoid explicit modeling and directly learn how to synthesize contrast-weighted images from the MRF data using neural networks. We refer to this approach as direct contrast synthesis (DCS).
In previous work, ~\cite{virtueetal} proposed a supervised DCS method in which a network was trained to take a single voxel MRF time series and map it to a specific contrast weighting (\emph{e.g.}, T1w, T2w, FLAIR). This approach, which we refer to as \textbf{PixelNet}, is illustrated in Figure~\ref{fig:dcscomparisons}b.
By training on many pairs of MRF and contrast-weighted images, PixelNet can achieve better results than dictionary mapping and simulation-based contrast synthesis. However, by processing each pixel independently, PixelNet does not leverage the spatial structure in the data and thus can suffer from noise and spatial inconsistency.
To address this issue, we propose to implement DCS as an image-to-image translation task to leverage structural information. In the field of computer vision, image-to-image translation is an established problem that aims to translate an image from a source domain to a target domain (e.g., reconstructing objects from edge maps~\cite{isola2017image} and colorizing images~\cite{zhang2016colorful}).
Recent studies, in particular, have shown promising results through image-to-image Convolutional Neural Networks (CNNs), and Generative Adversarial Networks (GANs)~\cite{goodfellow2020generative,isola2017image}.
The seminal work of pix2pix~\cite{isola2017image} investigates conditional adversarial networks as a general-purpose solution to image-to-image translation problems. 
CycleGAN~\cite{Zhu_2017_ICCV} improved upon the technique to learn image-to-image translation in the absence of paired examples. 
Image-to-image translation has also been applied in the field of medical imaging and MRI. For example, ~\cite{maspero2018dose,han2017mr,wolterink2017deep} learn cross-modality image synthesis between MRI and CT images; ~\cite{yu2019ea,dar2019image} synthesizes T2w images from T1w images; ~\cite{qu2020synthesized} synthesizes 7T high-resolution, high-SNR images from 3T input images; and ~\cite{wang2020synthesize} introduces a multi-task deep learning model to synthesize multi-contrast MRI images from multi-echo sequences.
Recently, ~\cite{dalmaz2022resvit} proposed a residual transformer-based deep learning model for multi-modal cross-contrast MR image synthesis.

Inspired by previous works, we propose to use a conditional GAN-based architecture for DCS from MRF and demonstrate substantial improvement in image quality and computation efficiency compared to simulation-based contrast synthesis and PixelNet.
We refer to our approach as \textbf{N-DCSNet}, first described in~\cite{wang2020high}, where $\mathbf{N}$ represents $N$ different contrasts that can be synthesized by our network (here $N=3$). Figure~\ref{fig:dcscomparisons} summarizes the three pipelines of producing synthetic, multi-contrast images.

As illustrated in Figures~\ref{fig:mrfoverview} and~\ref{fig:dcscomparisons}c, \textbf{N-DCSNet} directly synthesizes different contrast weighted images, i.e., T1w, T2w, FLAIR, from the MRF time series data through a spatial CNN. Our generator is designed as a U-Net with a single encoder and multi-branch decoders ~\cite{ronneberger2015u}. We implement a multi-layer CNN (PatchGAN)~\cite{isola2017image} as the discriminator.
The generator is based on spatial convolutions, allowing the network to learn and exploit spatial structural information.
Different contrast-weighted outputs share the same encoder to exploit the shared information across contrasts.
Separate decoders are designed to learn the unique features of each contrast.
During the training procedure, we leverage a conditional GAN framework, where the time average of the MRF time series is also used as an input to the discriminator to constrain the GAN training. 

\emph{In-vivo} experiments on healthy volunteers show that \textbf{N-DCSNet} can generate high-fidelity, multi-contrast images from MRF time-series. Our approach outperforms contrast synthesis from parameter maps and PixelNet both qualitatively and quantitatively. Furthermore, we demonstrate that \textbf{N-DCSNet} can inherently mitigate some artifacts that appear in MRF, such as slice in-flow artifacts and spiral off-resonance blurring.
Our main contributions can be summarized as follows:
\begin{itemize}
  \item We introduce a spatial CNN-based method to learn the mapping between MRF time series and contrast-weighted images (\emph{i.e.,} T1w, T2w, FLAIR). Our approach can avoid simulation errors typically seen in Synthetic MR.
  \item We use a conditional GAN-based framework to encourage finer textures and produce more faithful contrasts.  Additionally, our \textbf{N-DCSNet} can inherently mitigate slice in-flow artifacts as well as spiral off-resonance blurring.

  \item \textbf{N-DCSNet} outperforms simulation-based contrast synthesis from parameter maps and PixelNet both qualitatively as well as based on quantitative metrics. It also has significant computation advances.  During inference, our approach is significantly faster than simulation-based contrast synthesis and PixelNet, thus improving the potential for clinical adoption.
\end{itemize}

\section{Material and methods}

In this section, we first describe the data acquisition protocols and the simulation-based contrast synthesis via parameters used as our baseline for comparisons (\S~\ref{sec:data}). Then, we introduce our GAN-based framework design for \textbf{N-DCSNet} (\S~\ref{sec:gandesign}). 
Next, we detail the loss functions (\S~\ref{sec:loss}) and the training process. Finally, we compare our method with previous approaches (\S~\ref{sec:exp}).

\begin{figure}[!ht]
\hspace{-30mm}
\includegraphics[width=1.25\linewidth]{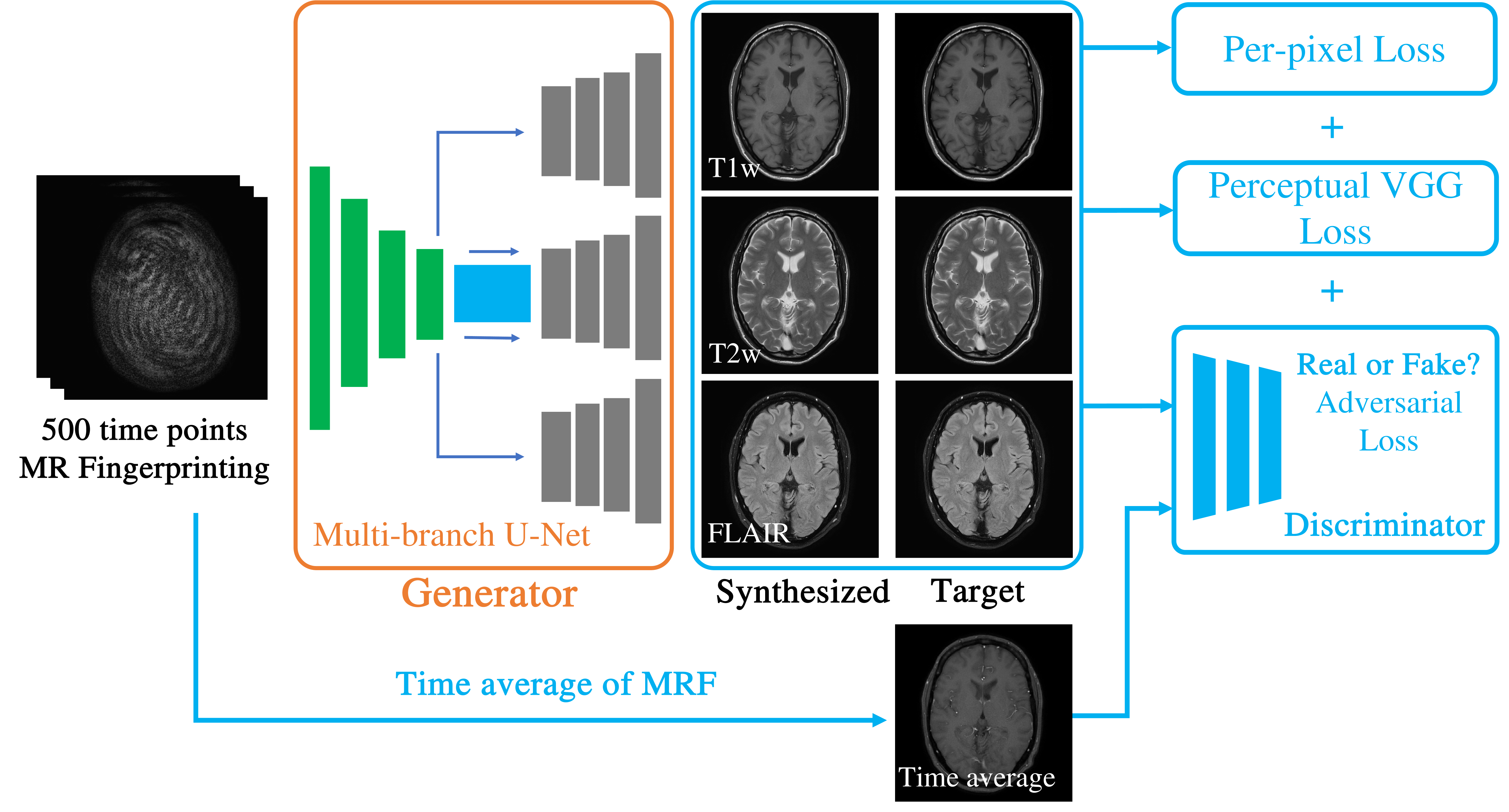}
\caption{\label{fig:overall}\textbf{Illustration of our proposed N-DCSNet framework.} Given a complex-valued MRF time series $\mathbf{MRF}_{in} \in \mathbb{C}^{t\times h\times w}$, with number of time points $t\in \mathbb{N}$ and image dimensions $h,w \in \mathbb{N}$, \textbf{N-DCSNet} synthesizes three contrast-weighted images (T1w, T2w, FLAIR) with a single network. We design a multi-branch U-Net as the generator and a multi-layer CNN as the discriminator following the conditional GAN training strategy. To constrain the GAN training, we additionally input the time average of MRF to the discriminator.
A combination of per-pixel $\ell_1$ loss, perceptual VGG loss, and adversarial loss is imposed on the network. \textbf{N-DCSNet} generates high-fidelity contrast-weighted images with sharper edges, finer textures, and more faithful contrasts compared to simulation-based contrast synthesis and PixelNet.  }
\end{figure} 

\subsection{Data acquisition and contrast synthesis via parameters\label{sec:data}}

\noindent\textbf{Data acquisition:} With IRB approval, we scanned 21 male subjects, ranging from 29 to 61 years old, with a 1.5T Philips Ingenia scanner using a 15-channel head coil. 13 channels were selected using automatic coil selection.
To avoid conducting so-called "data crimes"~\cite{shimron2022implicit}, here we report our data preparation pipeline as follows;
Four consecutive axial brain scans  were acquired for each exam session. The subjects were instructed to remain still throughout the exam so data across scans remains registered. 
\begin{itemize}
    \item A spoiled gradient echo (FISP)~\cite{jiang2015mr} MRF sequence with 500 time points, constant TE=3.3 ms and TR=20 ms. Each TR consisted of a spiral in-out readout where the spiral-out was rotated by 180 degrees with respect to the spiral-in. The spirals between two consecutive TR were rotated by 9 degrees. 
    \item T1w spin echo with TE=15 ms, TR=450 ms, flip angle=$69^{\circ}$ and two averages.
    \item T2w turbo spin echo (TSE) with TE=110 ms, TR=1990-2215 ms, ETL=16, flip angle=$90^{\circ}$ and two averages.
    \item FLAIR inversion recovery TSE with TE=120 ms, TR=8500 ms, TI=2500 ms, ETL=41, Flip angle=$90^{\circ}$ and two averages.

\end{itemize}

All scans were acquired with an in-plane resolution of 0.72$\times$0.72
mm (FOV 230$\times$230 mm, matrix size 320$\times$320) and nine to ten slices with a thickness of 5 mm.
Out of the 21 subjects, 17 were scanned twice (on different days), resulting in a total of 38 exams.
FLAIR sequences were acquired for only 26 of the 38 exams. Only the 26 exams with all four sequences were used in this study, out of which 21 were used for training, 2 for validation, and 3 for testing.
The data from subjects used for testing were not included in any of the training sets.

\noindent\textbf{Pre-processing:}
Each of the three contrast-weighted image data was normalized with respect to the 95th percentile of the intensity values for each image.
MRF time series images were reconstructed from each TR using gridding with density compensation~\cite{o1985fast,jackson1991selection} followed by coil combination with Philips' CLEAR. The MRF data was then normalized in the following way:  For each dataset, an averaged image from the 500 time points was computed. The 95th percentile of the magnitude values from the average MRF image was then used to normalize the time series.

\noindent\textbf{Parameter maps and contrast simulation:} The dictionary for MRF parameter mapping was simulated using EPG~\cite{weigel2015extended}.
The dictionary consisted of 22,031 MRF signals with T1 parameters ranging from 4 to ms-3,000 ms and T2 parameters ranging from 2 ms - 2,000 ms.
Each simulated signal in the dictionary was scaled to have a Euclidean norm equal to one.
We used cosine similarity~\cite{ma2013magnetic} to match the acquired MRF signal to the nearest neighbor in the simulated dictionary (Figure~\ref{fig:dcscomparisons} a).
Additional factors, such as B1 inhomogeneity and slice profile, were not included in the simulated dictionary.

The parameter maps (T1, T2) obtained from dictionary matching were then used to simulate the contrast-weighted images. The T1w spin echo (SE) has a closed form for specific TE and TR, and PD parameters:

\begin{equation}
    \mathrm{SE(PD, T1, T2, TE, TR)} = \mathrm{PD}\cdot (1-e^{-\frac{\mathrm{TR-TE}}{\mathrm{T1}}})\cdot e^{-\frac{\mathrm{TE}}{\mathrm{T2}}}.
    \label{eq:se}
\end{equation}
PD was computed by taking the magnitude of the inner product between the acquired MRF signal and the nearest neighbor in the simulated dictionary.
The T2w and FLAIR sequences are based on Turbo-Spin-Echo (TSE) and do not have closed forms. For these, we used EPG~\cite{weigel2015extended} to simulate the contrast-weighted images.

\subsection{\textbf{N-DCSNet} framework \label{sec:gandesign}}

Figure~\ref{fig:overall} illustrates the overall pipeline of our proposed \textbf{N-DCSNet}. Our network expects the complex-valued MRF time series $\mathbf{MRF}_{in} \in \mathbb{C}^{t\times h\times w}$ as input, where $t,h$, and $w$ correspond to the number of time points, image height, and image width, respectively ($t,h,w\in\mathbb{N}$). The network outputs are real-positive (magnitude) contrast weighted images $\hat{\textbf{T1w}},\hat{\textbf{T2w}},\hat{\textbf{FLAIR}} \in \mathbb{R}^{h\times w}$. 
 In our experiments, $t = 500, h = w = 320$.

We design a conditional GAN-based framework for \textbf{N-DCSNet},
the standard framework in ~\cite{isola2017image,Zhu_2017_ICCV}, consisting of a generator ($G$) and a discriminator ($D$). First, the real and imaginary parts of the MRF data are concatenated along the time dimension. 

Our generator is a modified U-Net~\cite{ronneberger2015u}, which consists of one shared encoder and multiple independent decoders.
The shared encoder exploits structural similarities across the multi-contrast images, whereas the independent decoders  learn the unique features of the different contrasts.
At test time, \textbf{N-DCSNet} produces multi-contrast images with a single network. 
The discriminator ($D$) is a multi-layer CNN (patchGAN)~\cite{isola2017image} that penalizes structure at a patch scale. $D$ aims to classify if each $N\times N$ patches in an image are real or fake.
We run this discriminator convolutionally across the image, averaging all responses to provide the final output of $D$.
To constrain the GAN training, we follow~\cite{isola2017image} and further input the magnitude of the MRF time-averaged image to the discriminator to provide structural guidance. This image has mixed contrast and, due to averaging, significantly reduced spiral undersampling artifacts compared to the MRF time-series images.
We denote it as $\mathbf{MRF}_{avg}$. 

During training, the generator $G$ learns to predict high-quality contrast-weighted images that cannot be distinguished from the real acquired images (ground truth) by an adversarially trained discriminator $D$.
Meanwhile, $D$ is simultaneously trained  to distinguish the generated images (labeled as "Fake") from the ground truth images (labeled as "Real").

\subsection{Loss functions \label{sec:loss}}

Our proposed \textbf{N-DCSNet} is fully supervised, with the purpose of generating high-fidelity contrast-weighted images that are close to the ground truth real acquisitions.
The loss function of our generator $G$ is a combination of three components: 1) $\ell_1$ reconstruction loss, 2) perceptual loss, and 3) adversarial loss.
%
Given our Generator $G$ and the input MRF signal $\mathbf{MRF}_{in}$, $G$ outputs the synthesized contrast-weighted images (T1w, T2w, FLAIR):

\begin{equation}
    \mathbf{\hat{T1w}}, \mathbf{\hat{T2w}}, \mathbf{\hat{FLAIR}} = G(\mathbf{MRF}_{in}).
    \label{eq:G}
\end{equation}
Then, the cumulative $\ell_1$ loss is formulated as:

\begin{equation}
    L_{\ell_1} = \mathbb{E}_{\mathbf{MRF}_{in}}(\lVert \mathbf{\hat{T1w}} - \mathbf{{T1w}} \rVert_{1} + \lVert \mathbf{\hat{T2w}} - \mathbf{{T2w}} \rVert_{1} + \lVert \mathbf{\hat{FLAIR}} - \mathbf{{FLAIR}} \rVert_{1}),
    \label{eq:l1}
\end{equation}
where $\mathbf{{T1w}},\mathbf{{T2w}},\mathbf{{FLAIR}}$ represent the real, ground-truth acquisitions of the three contrast-weighted images, respectively (\S~\ref{sec:data}).
Per-pixel losses such as the $\ell_1$ loss are known to exhibit image blurring ~\cite{isola2017image,johnson2016perceptual,larsen2016autoencoding,wang2022high}.
Therefore, we incorporate additional perceptual and adversarial losses to encourage detailed reconstructions.

Perceptual losses~\cite{johnson2016perceptual, wang2022high} have been used successfully in super-resolution and image synthesis~\cite{wang2020synthesize} tasks to improve image quality and encourage delicate structures. The idea is that layer features of task-based networks, like image classification networks, can capture high-level perceptual information in the image. Therefore, minimizing the loss in the feature space can preserve such perceptual information~\cite{johnson2016perceptual}.
In this work, the perceptual loss is implemented as the $\ell_2$ distance between \textbf{relu2-2} layer features of an ImageNet~\cite{deng2009imagenet} pre-trained VGG Network~\cite{simonyan2014very}.
We denote the function used to extract these features as $\phi(\cdot)$.
Each contrast-weighted image is scaled to $[0,1]$, duplicated three times, and concatenated along the channel dimension (to simulate RGB channels) before feeding into $\phi(\cdot)$. 
Then, the overall VGG perceptual loss term can be written as:

\begin{equation}
\hspace{-15mm}
    L_{\mathrm{vgg}} = \mathbb{E}_{\mathbf{MRF}_{in}}(\lVert \phi\mathbf{(\hat{T1w}})- \phi(\mathbf{{T1w})} \rVert_{2} + \lVert \phi(\mathbf{\hat{T2w}} )- \phi(\mathbf{{T2w}}) \rVert_{2} + \lVert \phi(\mathbf{\hat{FLAIR}}) - \phi(\mathbf{{FLAIR}}) \rVert_{2}).
    \label{eq:l1}
\end{equation}

The third component of our loss function is an adversarial loss. This term is used to further encourage high-frequency details and achieve more realistic synthesized outputs~\cite{isola2017image}.
The generator $G$ is trained to produce outputs that cannot be distinguished from “real” images.
We concatenate the acquired images $[\mathbf{MRF}_{avg},\mathbf{T1w},\mathbf{T2w},\mathbf{FLAIR}]$ along the channel dimension and treat it as the "real" sample $\mathbf{S}_{real} = [\mathbf{MRF}_{avg},\mathbf{T1w},\mathbf{T2w},\mathbf{FLAIR}]$.
Meanwhile, we create $\mathbf{S}_{fake} = [\mathbf{MRF}_{avg},\mathbf{\hat{T1w}},\mathbf{\hat{T2w}},\mathbf{\hat{FLAIR}}]$ as the "fake" sample $\mathbf{S}_{fake}$.
Then, the adversarial loss for our generator is given by:

\begin{equation}
    L_{\mathrm{adv}} = -\mathbb{E}_{\mathbf{S}_{fake}}[\log(D(\mathbf{S}_{fake}))].
    \label{eq:adg-G}
\end{equation}
The overall objective function for the generator becomes:
\begin{equation}
    L_{\mathrm{G}} = L_{\ell_1} + \lambda_{\mathrm{vgg}}L_{\mathrm{vgg}} + \lambda_{\mathrm{adv}}L_{\mathrm{adv}},
    \label{eq:all-G}
\end{equation}
where $\lambda_{\mathrm{vgg}}$ and $\lambda_{\mathrm{adv}}$ are the weights of the perceptual loss and adversarial loss, respectively.
In our experiments, we empirically set $\lambda_{\mathrm{vgg}}=0.03$ and  $\lambda_{\mathrm{adv}}=0.015$.

Our discriminator is adversarially trained to detect the generators' outputs as "fake" images. Following~\cite{goodfellow2020generative}, the objective function for our discriminator $L_{\mathrm{D}}$ is given by:

\begin{equation}
    L_{\mathrm{D}} = -\mathbb{E}_{\mathbf{S}_{real}}[\log(D(\mathbf{S}_{real}))] -\mathbb{E}_{\mathbf{S}_{fake}}[\log(1 - D(\mathbf{S}_{fake}))].
    \label{eq:adg-G}
\end{equation}

We update the parameter weights of $G$ and $D$ by alternatively minimizing the objectives $L_{\mathrm{G}}$ and $L_{\mathrm{D}}$.

\subsection{Experiments \label{sec:exp}}

To demonstrate its effectiveness, our \textbf{N-DCSNet} is evaluated against simulation-based contrast synthesis (Synthesis via parameters) and PixelNet on the same testing dataset (detailed in \S\ref{sec:data}).
%
%
The  EPG simulation using the dictionary-matched parameters was run for all voxels in parallel using the joblib package~\cite{joblib} on 24 CPUs.
Based on the architecture introduced in~\cite{virtueetal}, we implemented PixelNet as a 1-D temporal CNN to map the MRF time series at every voxel to the corresponding three contrast-weighted scans.
The PixelNet network consists of three convolutional layers followed by three fully-connected layers and is trained with an $\ell_2$ loss.
The inference time for the different approaches is calculated by computing the average runtime of 20 separate runs of a single MRF slice.
Ablations studies were also conducted to analyze the impacts of the different loss functions on the synthesized contrast-weighted images. 

\subsubsection{Evaluation metrics \label{sec:metric_eval}}
To quantitatively compare our results to the ground truth, we report the following evaluation metrics: normalized Root Mean Square Error (nRMSE), Peak Signal-to-Noise Ratio (PSNR), Structural
Similarity (SSIM)~\cite{wang2004image}, Learned Perceptual Image Patch Similarity (LPIPS)~\cite{zhang2018perceptual} with AlexNet~\cite{krizhevsky2017imagenet}, and Fréchet inception distance (FID) score~\cite{heusel2017gans}.
%
%
%
%
%
When computing LPIPS and FID, the output images were scaled to the range $[0,255]$ and saved as png files.

\subsubsection{Implementation details}

All the proposed algorithms and networks were implemented using PyTorch 1.8~\cite{paszke2019pytorch} on 24 GB NVIDIA 3090 graphics processing units (GPUs).
Our generator and discriminator were trained using Adam optimizer~\cite{kingma2014adam}, with a batch size of $4$, and a learning rate of $1\times 10^{-4}$.

We supervise the direct contrast synthesis with magnitude contrast weighted images. However, the MRF time series is inherently complex-valued. To reduce the sensitivity to phase, during training, we augmented the phase of the MRF data on the fly by multiplying each time-series with random constant phase $e^{j\theta}$, where $j = \sqrt{-1}$ and $\theta$ is uniformly distributed between $[0,2\pi]$.

%
%
%
The ablation study evaluating the loss functions contributions was performed by comparing the proposed combined loss function (Equation~\ref{eq:all-G}), against $L_{\ell_1}$, $L_{\ell_1} + \lambda_{\mathrm{vgg}}L_\mathrm{vgg}$, and $L_{\ell_1} + \lambda_{\mathrm{adv}}L_\mathrm{adv}$ losses.


\section{Results}

\subsection{Comparisons with contrast synthesis via parameters and PixelNet}

\begin{figure}[!ht]
\hspace{-25mm}
\includegraphics[width=1.2\linewidth]{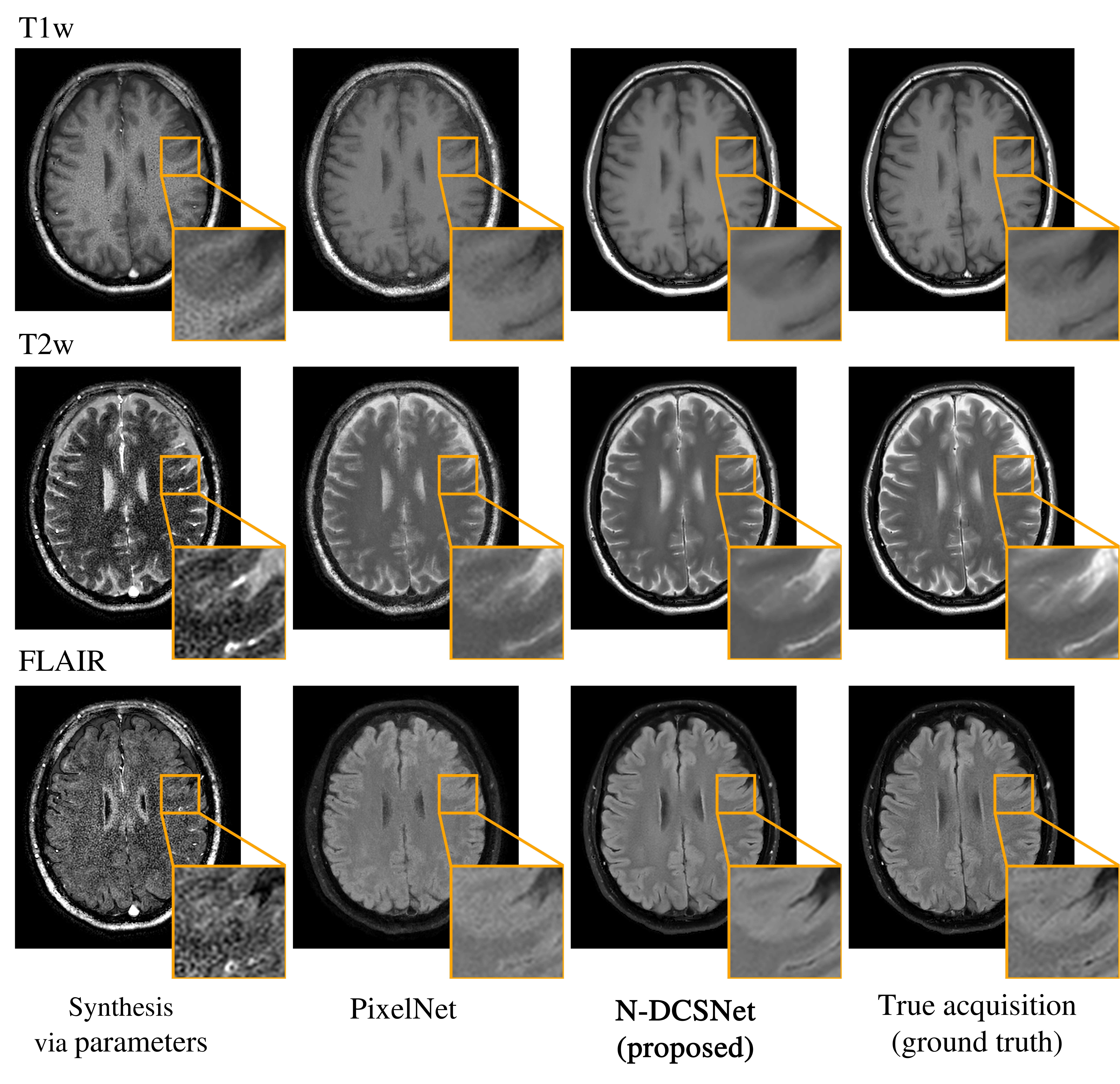}
\caption{\label{fig:results_1}\textbf{Representative contrast synthesis results of different methods (upper brain).} From left to the right, we compare our proposed \textbf{N-DCSNet} with simulation-based contrast synthesis via parameters~\cite{weigel2015extended}, PixelNet~\cite{virtueetal}, and the true acquisition. \textbf{N-DCSNet} shows better visual agreement with the true acquisition, producing finer textures and higher overall image quality compared to other approaches. Zoomed-in details are displayed next to each image.}

\end{figure}

Figure~\ref{fig:results_1} summarizes the results of the different contrast synthesis methods applied to a representative 2D brain slice.
Compared to EPG simulation-based synthesis (synthesis via parameters)~\cite{weigel2015extended}, and PixelNet~\cite{virtueetal}, \textbf{N-DCSNet} produces  finer  and cleaner structural details, sharper edges, and better perceptual agreement with the true acquisition (ground truth).
The EPG simulation-based results (synthesis via parameters) exhibit incorrect contrast and noise artifacts due to the modeling and acquisition imperfections (as expected in \S\ref{sec:introduction}).
PixelNet significantly improves the synthesized image quality, but the noise artifacts persist (as shown in T1w, and T2w).
In comparison, \textbf{N-DCSNet} leverages both temporal and spatial information, producing more faithful contrast, preserving finer details, and showing better agreements with the ground truth images.

\begin{figure}[!ht]
\hspace{-25mm}
\includegraphics[width=1.2\linewidth]{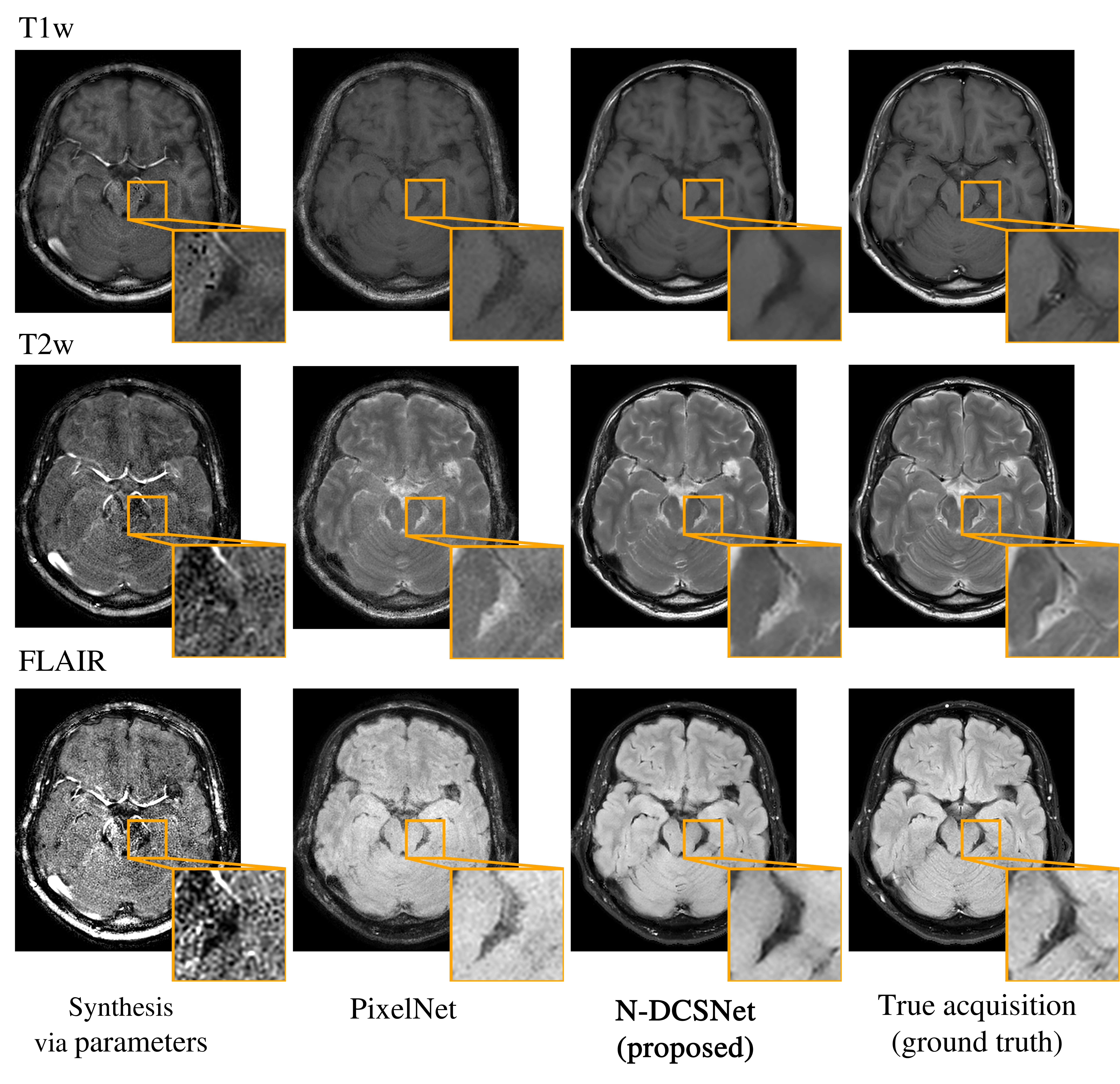}
\caption{\label{fig:results_2}\textbf{Representative contrast synthesis results of different methods (lower brain).} From left to the right, we compare our proposed \textbf{N-DCSNet} with simulation-based synthesis via parameters~\cite{weigel2015extended}, PixelNet~\cite{virtueetal}, and the true acquisition. Zoomed-in images point at the inflows (vasculature) region where parameter-based synthesis (left column) fails at delivering correct contrast due to the moving blood flow. In comparison, \textbf{N-DCSNet} can successfully reconstruct delicate textures and produce high-quality contrast-weighted images.}

\end{figure} 

Figure~\ref{fig:results_2} compares the results of another representative 2D slice from the lower brain.
Regions of the vasculature are zoomed in and expanded at the bottom right corners. Due to the blood flow, MRF cannot retrieve accurate parameter maps by dictionary matching~\cite{flassbeck2019flow}.
Therefore, synthesis via parameters fails to deliver precise contrasts in the vasculature regions (as shown in T2w images).
In contrast, \textbf{N-DCSNet} produces accurate contrast and can successfully reconstruct the delicate vessel structures (as shown in T2w results).
From the synthesized FLAIR images, we observe that PixelNet produces noisier images with flattened contrasts in the back of the brain. Instead, \textbf{N-DCSNet} successfully depicts the detailed textures and produces high-quality, sharper images.

Table~\ref{table:table_metric} compiles the quantitative evaluation metrics (nRMSE, PSNR, SSIM, LPIPS, FID) of different methods (Synthesis via parameters~\cite{weigel2015extended}, PixelNet~\cite{virtueetal} and \textbf{N-DCSNet}) for each contrast.
We compute the metrics across the testing dataset and report the mean and standard deviation (std).
As indicated in the table, for all three contrasts (T1w, T2w, FLAIR), our method consistently outperforms other methods in all five evaluation metrics. 
It is worth noting that LPIPS and FID use learned features to measure perceptual similarity between two images or two distributions, which has been shown to match better with human judgment than pixel-wise (nRMSE) or patch-wise (SSIM) metrics~\cite{zhang2018perceptual,heusel2017gans}. 
\textbf{N-DCSNet} significantly reduces the LPIPS by over 30\% and the FID by over 50\% compared to PixelNet for all three contrasts, demonstrating the superiority of our proposed method in terms of perceptual image quality.
\begin{table}[!ht]
\hspace{-28mm}
\centering
\setlength\tabcolsep{4.5pt} 
\begin{tabular}{l|l|lllll}
Contrasts              & Methods                                                           & nRMSE (\%) $\downarrow$ & PSNR (dB) $\uparrow$ & SSIM $\uparrow$& \begin{tabular}[c]{@{}l@{}}LPIPS $\downarrow$\\ ($\times 10^{-2}$)\end{tabular} & FID $\downarrow$  \\ 
\hline
\multirow{3}{*}{T1w}   & \begin{tabular}[c]{@{}l@{}}Synthesis\\via parameters ~\cite{weigel2015extended}\end{tabular} &  6.44 $\pm$ 1.25    &   24.0 $\pm$ 1.93    &  0.786 $\pm$ 0.030    &  20.1 $\pm$ 1.14     &   130.8   \\
                       & PixelNet~\cite{virtueetal}                                                        &   4.58 $\pm$ 0.83    &  26.9 $\pm$ 1.71   &  0.880 $\pm$ 0.026    &  11.3 $\pm$ 1.81     &  109.6    \\
                       & N-DCSNet (Ours) &  \colorbox{graynew}{\textbf{3.57 $\pm$ 0.64}}   &    \colorbox{graynew}{\textbf{29.1 $\pm$ 1.63}}  &  \colorbox{graynew}{\textbf{0.923 $\pm$ 0.019}}    &  \colorbox{graynew}{\textbf{6.33 $\pm$ 1.87}} &   \colorbox{graynew}{\textbf{57.32}}   \\ 
\hline
\multirow{3}{*}{T2w}   & \begin{tabular}[c]{@{}l@{}}Synthesis\\via parameters~\cite{weigel2015extended} \end{tabular} &    13.4 $\pm$ 1.68     &  17.5 $\pm$ 1.11    &   0.671 $\pm$ 0.032     &    21.1 $\pm$ 1.60   &   148.1   \\
                       & PixelNet~\cite{virtueetal}                                                          &   5.24 $\pm$ 0.64   &   25.7 $\pm$ 1.11   &   0.853 $\pm$ 0.027    &  12.6 $\pm$ 1.92     &    114.1  \\
                       & N-DCSNet (Ours)                                                         &  \colorbox{graynew}{\textbf{3.76 $\pm$ 0.59}}     &  \colorbox{graynew}{\textbf{28.6 $\pm$ 1.35}}    & \colorbox{graynew}{\textbf{0.921 $\pm$ 0.017}}     &   \colorbox{graynew}{\textbf{5.77 $\pm$ 1.02}}      & \colorbox{graynew}{\textbf{57.01}}     \\ 
\hline
\multirow{3}{*}{FLAIR} & \begin{tabular}[c]{@{}l@{}}Synthesis\\via parameters~\cite{weigel2015extended}\end{tabular} &   19.4 $\pm$ 2.75    &  14.3 $\pm$ 1.25    &   0.576 $\pm$ 0.028   &  20.6 $\pm$ 2.50     &  185.4    \\
                       & PixelNet~\cite{virtueetal}                                                          &  4.69 $\pm$ 0.67     & 26.7 $\pm$ 1.30     &  0.797 $\pm$ 0.025    &   11.3 $\pm$ 1.35  &   126.9   \\
                       & N-DCSNet (Ours)                                                           &   \colorbox{graynew}{\textbf{3.64  $\pm$ 0.65}}  &   \colorbox{graynew}{{\textbf{29.0 $\pm$ 1.75}}}  &  \colorbox{graynew}{{\textbf{0.883 $\pm$ 0.018}} }   &    \colorbox{graynew}{\textbf{8.63 $\pm$ 0.839}}   &    \colorbox{graynew}{\textbf{63.17}}  \\
\hline
\end{tabular}
\vspace{4mm}

\caption{\label{table:table_metric}\textbf{Quantitative comparisons (nRMSE, PSNR, SSIM, LPIPS, FID) between different contrast synthesis methods (mean $\pm$ standard deviation).} We calculate the metrics for each contrast (T1w, T2w, FLAIR) separately. \textbf{N-DCSNet} is compared with contrast synthesis via parameters~\cite{weigel2015extended}, and PixelNet~\cite{virtueetal}. Our proposed method consistently outperforms other approaches in all five metrics for each contrast. \colorbox{graynew}{\textbf{Bold}} corresponds to the best results. $\uparrow$ means higher the better, $\downarrow$ means lower the better.}
\end{table}

Table~\ref{table:computation} summarizes the inference time of the different approaches.
As indicated in the table, simulation-based synthesis (synthesis via parameters) takes an average of $24.37$ seconds due to the time-consuming dictionary matching and contrast simulation procedures that are repeated for each voxel across the entire image.
PixelNet is more efficient and averages $0.3421$ seconds by leveraging  parallel computing on GPU(on a single NVDIA 3090).
In comparison, our \textbf{N-DCSNet} has 20 times faster inference time than PixelNet.
\textbf{N-DCSNet} takes an average of 0.01617 seconds to synthesize three contrast-weighted images from a single 2D MRF time series, demonstrating superior computation efficiency and the potential for clinical translation.
All the experiments were run on a single NVIDIA 3090 GPU.

\begin{table}
\centering
\begin{tabular}{l|lll} 
\hline
             & \begin{tabular}[c]{@{}l@{}}Synthesis~\\via parameters\end{tabular} & PixelNet & N-DCSNet (Ours)  \\ 
\hline
Inference time (s) $\downarrow$ &     24.37        &  0.3421        &   \colorbox{graynew}{\textbf{ {0.01617}}}       \\
\hline
\end{tabular}
\vspace{4mm}
\caption{\label{table:computation} \textbf{Inference time of different methods for contrast synthesis from a 2D MRF time series.} \textbf{N-DCSNet} significantly reduces the inference time by over 20 times compared to PixelNet, demonstrating superior computation efficiency and the potential for clinical adoption. All the experiments are implemented on a single NVIDIA 3090 GPU for fair comparisons. \colorbox{graynew}{\textbf{Bold}} corresponds to the best result.}
\end{table}

\subsection{Ablation study of the different loss functions}

\begin{figure}[!ht]
\hspace{-25mm}
\includegraphics[width=1.2\linewidth]{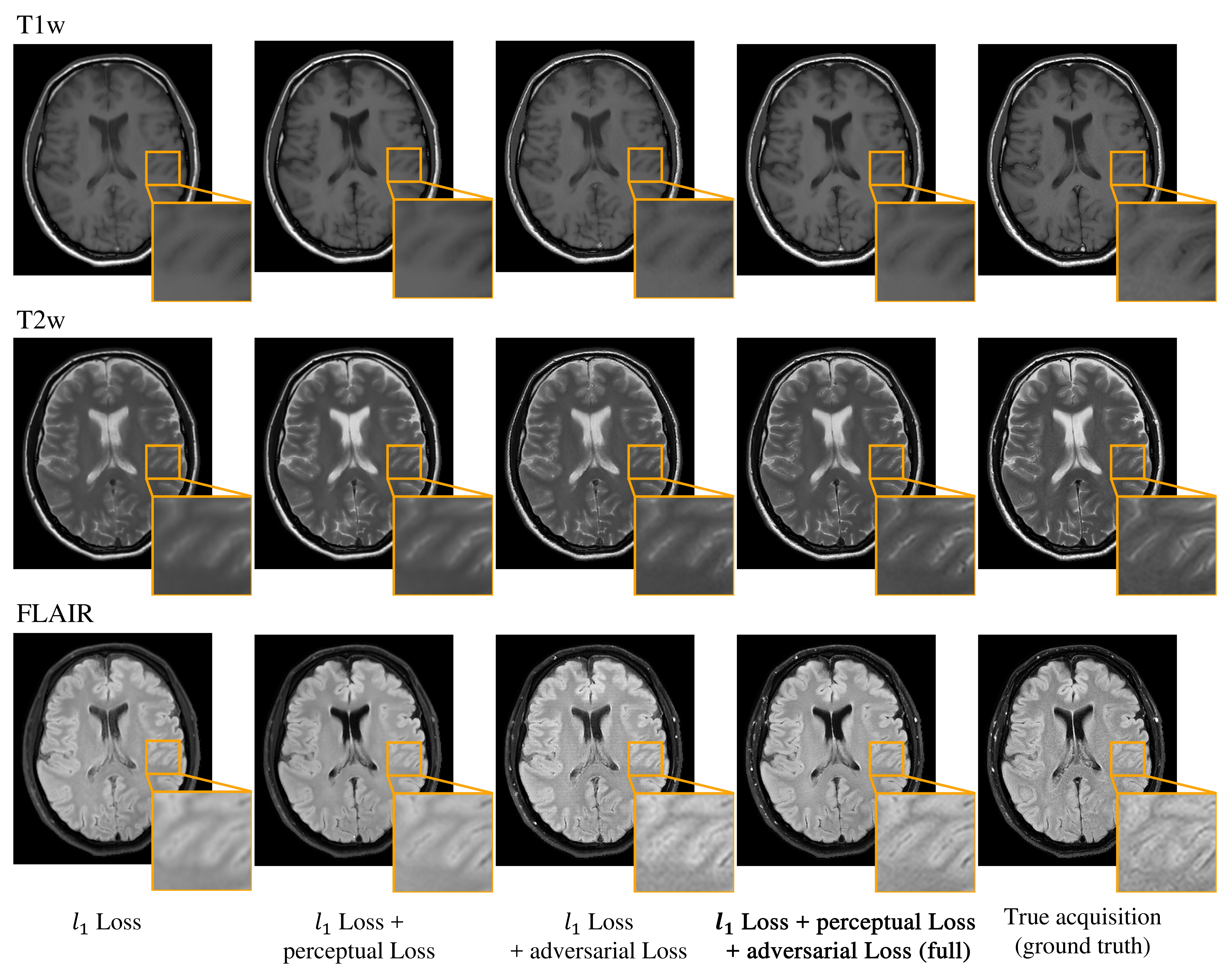}
\caption{\label{fig:ablation}\textbf{Representative visual comparison of N-DCSNet with different loss functions.} From left to right, we compare our full objective (fourth column - Equation \ref{eq:all-G}) with $L_{\ell_1}$, $L_{\ell_1} + \lambda_{\mathrm{vgg}}L_{\mathrm{vgg}}$, $L_{\ell_1} + \lambda_{\mathrm{adv}}L_{\mathrm{adv}}$ and the ground truth. Perceptual VGG loss encourages sharper edges compared to pure $L_{\ell_1}$, while adversarial loss further improves the image quality. The model trained with our full objective is able to recover subtle structures and show better visual agreements with the ground truth.}
\end{figure}

To investigate and better understand the impact of loss functions (\S~\ref{sec:loss}) on the resulting image quality,
we conducted an ablation study by comparing our overall loss function $L_\mathrm{G}$ (Equation~\ref{eq:all-G}) to $L_{\ell_1}$, $L_{\ell_1}+\lambda_{\mathrm{vgg}}L_{\mathrm{vgg}}$ and $L_{\ell_1}+\lambda_{\mathrm{adv}}L_{\mathrm{adv}}$ losses. We trained separate models with different objective functions using the same training setup and datasets (i.e., training set, learning rate, epochs, etc.).
Figure~\ref{fig:ablation} shows the results on a representative 2D brain slice. 
The model trained with pure $L_{\ell_1}$ (left column) suffers from degraded perceptual image quality and exhibits some blurring, which is consistent with the findings in literature~\cite{wang2022high,isola2017image,wang2020synthesize}.
Adding perceptual VGG loss (second column) encourages finer details and sharper edges. However, blurring artifacts remain (as seen in T2w and FLAIR).
Adding adversarial loss on top of $L_{\ell_1}$ (third column) encourages even finer structures but suffers from residual blurring (T2w) and recurrent checkerboard artifacts (FLAIR).  
By incorporating both perceptual loss and adversarial loss, the model trained with our proposed objective (fourth column, Equation \ref{eq:all-G}) can further improve the synthesized image quality, reconstructing more delicate textures (T2w example) and producing more faithful contrast (See the FLAIR example).

Table~\ref{table:table_metric_abl} summarizes the five evaluation metrics for \textbf{N-DCSNet} trained with different loss functions.
Since the model trained with pure $L_{\ell_1}$ loss optimizes the pixel distances, it produces the best nRMSE and PSNR results. 
However, it is known that nRMSE and PSNR do not match human perception~\cite{zhang2018perceptual}.
For perception-representative metrics (SSIM, LPIPS, FID), \textbf{N-DCSNet} trained with our proposed full objective outperforms the other loss functions for all three contrasts (except SSIM for T1w), which demonstrates the effectiveness of our loss functions in producing high-fidelity contrast-weighted images.

\begin{table}[!ht]
\hspace{-45mm}
\centering
\setlength\tabcolsep{4.5pt} 
\begin{tabular}{l|l|lllll}
Contrasts              & Methods                                                           & nRMSE (\%) $\downarrow$ & PSNR (dB) $\uparrow$ & SSIM $\uparrow$& \begin{tabular}[c]{@{}l@{}}LPIPS $\downarrow$\\ ($\times 10^{-2}$)\end{tabular} & FID $\downarrow$  \\ 
\hline
\multirow{3}{*}{T1w}   & $L_{\ell_1}$&  \colorbox{graynew}{\textbf{3.34 $\pm$ 0.63}}     & \colorbox{graynew}{\textbf{29.7 $\pm$ 1.69}}      &   {0.918 $\pm$ 0.018}   &   {8.02 $\pm$ 2.40}    &   67.39   \\

                       & $L_{\ell_1} + \lambda_{\mathrm{vgg}}L_{\mathrm{vgg}}$    &  3.43 $\pm$ 0.82     &   29.5 $\pm$ 2.16   & \colorbox{graynew}{\textbf{0.926 $\pm$ 0.022}}     &  {9.14 $\pm$ 2.53}     &  66.66    \\

                       & $L_{\ell_1} + \lambda_{\mathrm{adv}}L_{\mathrm{adv}}$    &  3.68 $\pm$ 0.93     &   28.7 $\pm$ 1.72   & {0.921 $\pm$ 0.020}     &  {8.04 $\pm$ 2.18}     &  62.94    \\
                       & $L_{\ell_1} + \lambda_{\mathrm{vgg}}L_{\mathrm{vgg}} + \lambda_{\mathrm{adv}}L_{\mathrm{adv}}$ (full) & {3.57 $\pm$ 0.64}   &    {29.1 $\pm$ 1.63}  &  {0.923 $\pm$ 0.019}    &  \colorbox{graynew}{\textbf{6.33 $\pm$ 1.87}} &   \colorbox{graynew}{\textbf{57.32}}   \\
\hline
\multirow{3}{*}{T2w}   &$L_{\ell_1}$&  \colorbox{graynew}{\textbf{3.57 $\pm$ 0.67}}    &  \colorbox{graynew}{\textbf{29.2 $\pm$ 1.64 }}   &  {0.914 $\pm$ 0.018}    & {10.08 $\pm$ 1.73}      &   71.44   \\
                       & $L_{\ell_1} + \lambda_{\mathrm{vgg}}L_{\mathrm{vgg}}$ & 3.67 $\pm$ 0.61 &   28.8 $\pm$ 1.45   & {0.918 $\pm$ 0.018}     &   {8.67 $\pm$ 1.46}    &    64.55 \\

                       & $L_{\ell_1} + \lambda_{\mathrm{adv}}L_{\mathrm{adv}}$    &  3.79 $\pm$ 0.72     &   28.4 $\pm$ 1.55   & {0.919 $\pm$ 0.024}     &  {7.57 $\pm$ 1.18}     &  60.30    \\                       
                       & $L_{\ell_1} + \lambda_{\mathrm{vgg}}L_{\mathrm{vgg}} + \lambda_{\mathrm{adv}}L_{\mathrm{adv}}$ (full)

                       &  {3.76 $\pm$ 0.59}     &  {28.6 $\pm$ 1.35}    & \colorbox{graynew}{\textbf{0.921 $\pm$ 0.017}}     &   \colorbox{graynew}{\textbf{5.77 $\pm$ 1.02}}      & \colorbox{graynew}{\textbf{57.01}}     \\ 
\hline
\multirow{3}{*}{FLAIR} & $L_{\ell_1}$& \colorbox{graynew}{\textbf{ 3.44 $\pm$ 0.66}}     & \colorbox{graynew}{\textbf{29.4 $\pm$ 1.72 }}    &  {0.879 $\pm$ 0.017}    & {11.1 $\pm$ 0.98}      &  93.01    \\
                       & $L_{\ell_1} + \lambda_{\mathrm{vgg}}L_{\mathrm{vgg}}$   &   3.73 $\pm$ 0.61    &  28.7 $\pm$ 1.55     & 0.878 $\pm$ 0.019     &    {10.7 $\pm$ 1.02}   & 96.01     \\
                       & $L_{\ell_1} + \lambda_{\mathrm{adv}}L_{\mathrm{adv}}$    &  3.68 $\pm$ 0.93     &   28.1 $\pm$ 1.71   & {0.869 $\pm$ 0.021}     &  {9.62 $\pm$ 1.08}     &  78.71    \\    
                       & $L_{\ell_1} + \lambda_{\mathrm{vgg}}L_{\mathrm{vgg}} + \lambda_{\mathrm{adv}}L_{\mathrm{adv}}$ (full)                                                             &   {3.64  $\pm$ 0.65}  &   {29.0 $\pm$ 1.75}  &  \colorbox{graynew}{\textbf{{0.883 $\pm$ 0.018}}}    &    \colorbox{graynew}{\textbf{{8.63 $\pm$ 0.839}}}   &    \colorbox{graynew}{\textbf{63.17}}  \\
\hline
\end{tabular}
\vspace{4mm}
\caption{\label{table:table_metric_abl}\textbf{Quantitative comparisons (nRMSE, PSNR, SSIM, LPIPS, FID) of N-DCSNet with different loss function designs (mean $\pm$ standard deviation). } The model trained with pure $L_{\ell_1}$ optimizes the per-pixel distances, producing the lowest nRMSE and highest PSNR. The model trained with our full objective outperforms other loss function designs in perceptual metrics SSIM, LPIPS, and FID.  \colorbox{graynew}{\textbf{Bold}} corresponds to the best results. $\uparrow$ means higher the better, $\downarrow$ means lower the better.}
\end{table}

\subsection{Mitigation of spiral off-resonance artifacts \label{mitigation}}

Besides the aforementioned superior performance, we also demonstrate cases where \textbf{N-DCSNet} can effectively mitigate the off-resonance artifacts within the MRF time series caused by B0 inhomogeneity and the long readout time of spiral acquisitions. 
Previous works have demonstrated the feasibility and potential of deep learning in off-resonance corrections~\cite{zeng2019deep,alfredo}. As shown in Figure~\ref{fig:results_1}, \ref{fig:results_2}, parameter-based synthesis and PixelNet present blurry scalp fat signals in boundary regions of the brain due to the MRF off-resonance effects (seen in T1w).
In comparison, benefiting from spatial convolutions, \textbf{N-DCSNet} reconstructs clean and sharp scalp fat signal, overcomes the off-resonance artifacts, and agrees well with the ground truth. 

\begin{figure}[!ht]
\hspace{-25mm}
\includegraphics[width=1.2\linewidth]{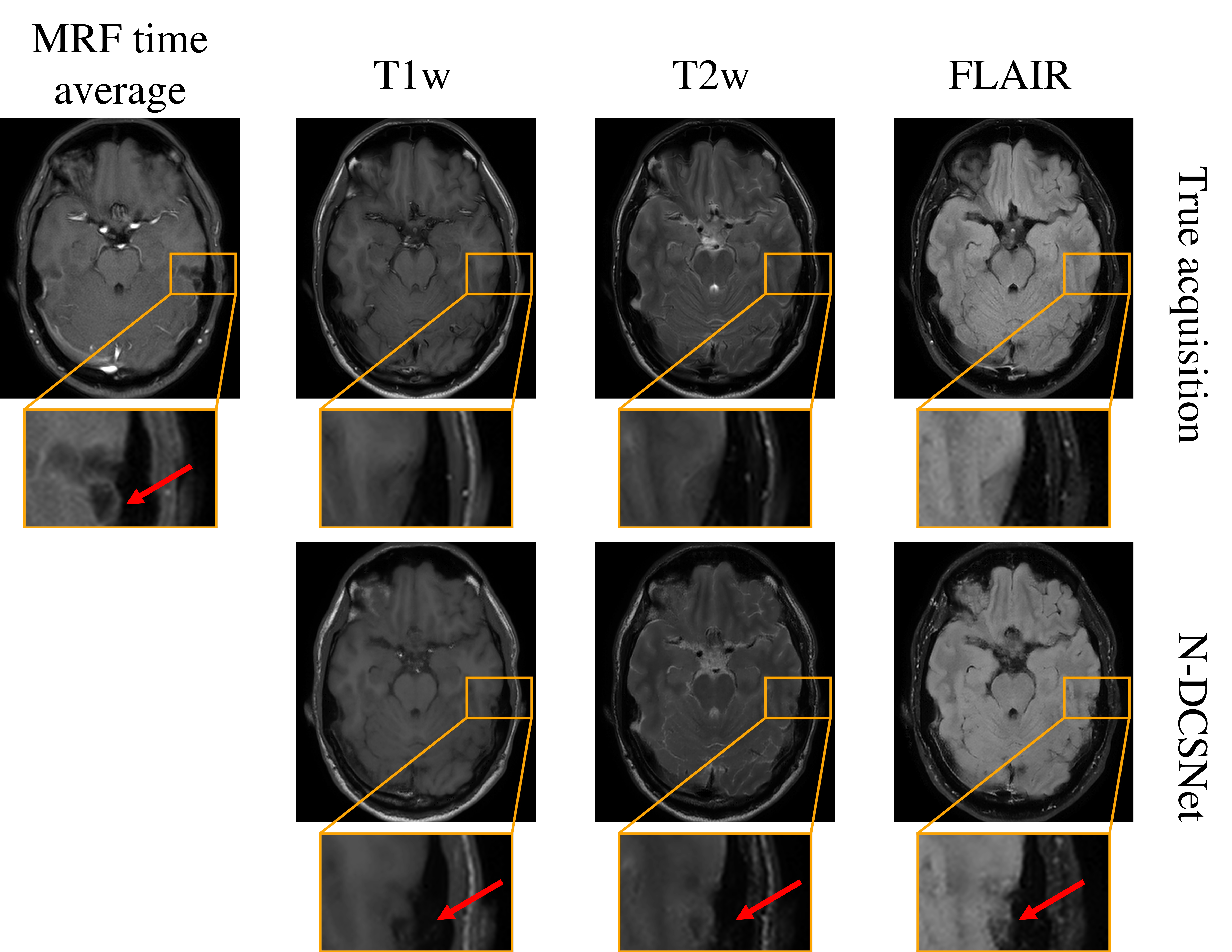}
\caption{\label{fig:off_resonance_1} \textbf{Representative N-DCSNet results in  mitigating off-resonance artifacts in MRF time series near the skull region.} MRF time-averaged image exhibits spiral off-resonance artifacts near the skull region (zoomed-in images) due to the long readout time. \textbf{N-DCSNet} can successfully mitigate the artifacts and produce contrast-weighted images with little residual artifacts. True acquisitions are displayed as references. Red arrows point at the regions with residual artifacts.}

\end{figure}

\begin{figure}[!ht]
\hspace{-25mm}
\includegraphics[width=1.2\linewidth]{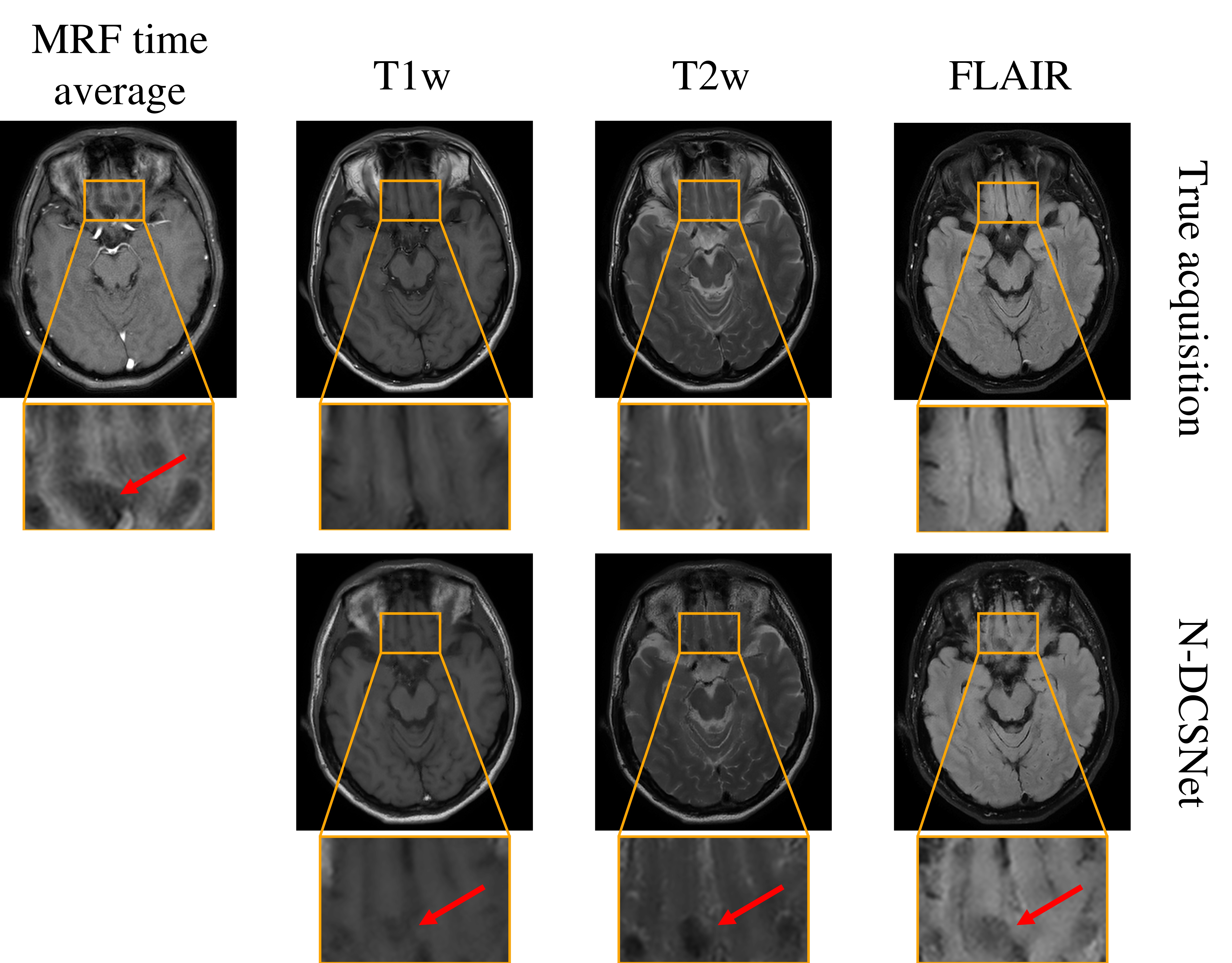}
\caption{\label{fig:off_resonance_2} \textbf{Representative N-DCSNet results in  mitigating spiral off-resonance artifacts in MRF time series near the nasal region.} MRF time-averaged image exhibits severe spiral off-resonance artifacts near the nasal region (zoomed-in images) due to B0 inhomogeneity and the long readout time. \textbf{N-DCSNet} can recover the structure and produce contrast-weighted images with little residual artifacts. True acquisitions are displayed as references. Red arrows point at the regions with residual artifacts.}

\end{figure}

Figure~\ref{fig:off_resonance_1} shows a representative example where the MRF time-averaged image exhibits off-resonance signal loss artifacts in the regions close to the skull (indicated by the zoomed-in details).
\textbf{N-DCSNet} significantly reduces the artifacts and recovers the correct contrasts and structures. Some residual artifacts can be observed, as indicated by the red arrows.
Figure~\ref{fig:off_resonance_2} presents another example where the MRF time-averaged image exhibits several off-resonance signal loss artifacts and geometric distortion near the nasal region. Most brain structures are blurred out. This is primarily due to the considerable B0 homogeneity and the long readout time for the spiral acquisition.
As visualized in the figure, \textbf{N-DCSNet} accurately recovers most of the delicate brain structures near the nasal region. The red arrows indicate the residual artifacts.

\section{Discussion}

In this work, we present a novel high-fidelity direct contrast synthesis framework \textbf{N-DCSNet} for synthesizing multi-contrast images from a single MRF scan. 
\textbf{N-DCSNet} directly learns a mapping between the MRF time series and the desired contrast weighted images (\emph{i.e.}, T1w, T2w, FLAIR) and thus bypasses the mapping and simulation steps required for contrast synthesis from parameter maps.

As briefly introduced in \S~\ref{sec:introduction}, the sources of error contrast synthesis via parameter maps is mainly attributed to: 1) factors that are not included in the dictionary simulation (e.g., B0/B1 homogeneity, slice profile, flow effects), 2) approximation and error propagation in the contrast synthesis simulation (EPG algorithm)~\cite{weigel2015extended}, 3) artifacts (noise and aliasing) from highly undersampled MRF scans (example shown in Figure \ref{fig:mrfoverview}). As indicated by the visual comparison results (Figure~\ref{fig:results_1},~\ref{fig:results_2}), the parameter-based contrast synthesis method fails to deliver the correct contrast and produces noisier outputs (especially for T2w and FLAIR results). One possible way to improve the results is by modeling more parameters during the dictionary simulation procedure, such as B1 inhomogeneity~\cite{buonincontri2016mr}, flow~\cite{flassbeck2019flow}, partial volume~\cite{deshmane2019partial}. Unfortunately, including more simulation parameters, forces the dictionary to grow in size, which prolongs the dictionary matching time (Table \ref{table:computation}), or severely sacrifices parameter resolution and range.

Direct contrast synthesis, or DCS, leverages paired training data to learn a mapping from MRF signals to contrast-weighted images without explicitly modeling the aforementioned conditions. 
Previous DCS method PixelNet~\cite{virtueetal} proposed a 1-D temporal CNN that maps the MRF time series at each pixel to the contrast weightings for that pixel and improves synthesized image quality and inference time (Tabel \ref{table:computation}). 
However, because PixelNet treats each pixel independently, it doesn't leverage the unique spatial structural information within the MRF data. 
\emph{In-{vivo}} results (Figure~\ref{fig:results_1},~\ref{fig:results_2}) indicate that PixelNet exhibits severe noise artifacts and diminished fine textures, especially in FLAIR scans.

Our \textbf{N-DCSNet} shows significant improvements by introducing a conditional GAN-based framework with a spatial convolution network as the generator. \textbf{N-DCSNet} produces more faithful contrasts and is able to recover finer structures with overall better image quality (Figure~\ref{fig:results_1} and ~\ref{fig:results_2}). 
Moreover, as described in section \S~\ref{mitigation} and shown in Figure~\ref{fig:off_resonance_1} and~\ref{fig:off_resonance_2}, we demonstrate cases where \textbf{N-DCSNet} can effectively mitigate spiral off-resonance artifacts.

One limitation of this work is that DCS frameworks (PixelNet and \textbf{N-DCSNet}) can only generate contrast-weighted images with fixed sequence parameters (e.g., TE, TR), which is less flexible compared to simulation-based contrast synthesis from parameter maps. Separate networks need to be trained for different MRF parameters or contrast acquisitions. Another limitation for our \textbf{N-DCSNet} is the need for pared data. However, our approach makes it possible to train each decoder branch separately, thus relaxing this constraint, though this requires further investigations. Additionally, in this work, we only trained \textbf{N-DCSNet} on a limited number of healthy volunteer data (21 exams, 203 slices). Larger and more diverse clinical training data (\emph{e.g.}, with pathology) is needed for future clinical adoptions.

In the future, we would like to extend our framework to more diverse contrast synthesis, which includes but is not limited to gradient echo imaging, diffusion-weighted imaging, and susceptibility-weighted imaging.


\section{Conclusion}
In this work, we propose \textbf{N-DCSNet} to directly synthesize multi-contrast MR images from a single MRF acquisition, which can significantly reduce exam time. 
By directly training a network to generate contrast-weighted images from MRF, our method does not require any model-based simulation and, therefore, can avoid reconstruction errors due to simulation. 
\emph{In-{vivo}} experiments demonstrate that \textbf{N-DCSNet} produces high-fidelity contrast-weighted images with sharper contrast and minimal artifacts (in-flow and spiral off-resonance artifacts), which significantly outperforms simulation-based contrast synthesis and PixelNet, both visually and metric-wise. Additionally, our proposed method can inherently mitigate some off-resonance artifacts within the MRF data, producing high-quality contrast-weighted images with minimal residual artifacts.

\vspace{10mm}

\noindent \textbf{Acknowledgments:} The authors thank Dr. Patrick M. Virtue for his helpful discussions and Yuhan Wen for her help with paper editing.

\vspace{4mm}

\noindent \textbf{Author Contributions:} Conceptualization, K.W., M.D., J.T., and M.L.; Methodology, K.W., M.D., J.M., T.A., E.K., F.T., and M.L.; Resources, M.D., J.M., T.A., J.T., S.Y., and M.L.; Data Curation, M.D., J.M., and T.A.; Writing – Original Draft Preparation, K.W.; Writing – Review \& Editing, K.W., M.D., J.M., E.K., J.T., and M.L.; Supervision, M.D., J.M., J.T., S.Y., and M.L.; 

\vspace{4mm}

\noindent \textbf{Institutional Review Board Statement:} The study was conducted in accordance with the Declaration of Helsinki, and approved by the Institutional Review Board of Philips Healthcare (protocol code: Bet -Regelung 01-09 MR+Felder; date of approval: November 24th, 2009).

\vspace{4mm}

\noindent \textbf{Informed Consent Statement:} Informed consent was obtained from all subjects involved in the study.
\vspace{4mm}

\noindent \textbf{Data Availability Statement:} Our source code and pre-trained models will be publicly released upon the publication of this manuscript.

\vspace{4mm}

\noindent \textbf{Conflicts of Interest:} M.D., J.M., and T.A. are employees of Philips Research Europe.

\begin{adjustwidth}{-\extralength}{0cm}

\clearpage
\newpage
\reftitle{References}




\bibliography{ref}

\end{adjustwidth}
\end{document}